\documentclass[traditabstract]{aa} 

\usepackage{txfonts}
\usepackage{natbib}                             
\usepackage{graphicx}
\usepackage{lscape}
\usepackage{url}
\usepackage{longtable}

\def\ltr{$L_{\rm X}-T$ }

\begin{document}
   \title{The 2XMMi/SDSS Galaxy Cluster Survey}
      \subtitle{III. Clusters associated with spectroscopically targeted luminous red 
galaxies in SDSS-DR10}

\titlerunning{The 2XMMi/SDSS Galaxy Cluster Survey. III. }

   \author{A. Takey \inst{1,2}, 
           A. Schwope\inst{1}, \and
           G. Lamer\inst{1}
          }

   \institute{Leibniz-Institut f{\"u}r Astrophysik Potsdam (AIP),
              An der Sternwarte 16, 14482 Potsdam, Germany\\
              \email{atakey@aip.de}
         \and
             National Research Institute of Astronomy and Geophysics (NRIAG), 
             11421 Helwan, Cairo, Egypt             
             }

   \date{Received ....; accepted ....}


\abstract
{
We present a sample of 383 X-ray selected galaxy groups and clusters with 
spectroscopic redshift measurements (up to $z \sim 0.79$) from the 
2XMMi/SDSS Galaxy Cluster Survey. The X-ray cluster candidates were selected 
as serendipitously detected sources from the 2XMMi-DR3 catalogue that were  
located in the footprint of the Sloan Digital Sky Survey (SDSS-DR7). 
The cluster galaxies with available spectroscopic redshifts were selected 
from the SDSS-DR10. We developed an algorithm for identifying the cluster 
candidates that are associated with spectroscopically targeted luminous red 
galaxies and for constraining the cluster spectroscopic redshift.
A cross-correlation of the constructed cluster sample with published optically 
selected cluster catalogues yielded 264 systems with available redshifts. 
The present redshift measurements are consistent with the published values.
The current cluster sample extends the optically confirmed cluster sample 
from our cluster survey by 67 objects. Moreover, it provides spectroscopic 
confirmation for 78 clusters among our published cluster sample, which previously 
had only photometric redshifts. Of the new cluster sample that comprises 
67 systems, 55 objects are newly X-ray discovered clusters and 52 systems are 
sources newly discovered as galaxy clusters in optical and X-ray wavelengths. 
Based on the measured redshifts and the fluxes given in the 2XMMi-DR3 catalogue, 
we estimated the X-ray luminosities and masses of the cluster sample. 
}


\maketitle


\section{Introduction}

Galaxy clusters are the largest gravitationally bound objects in the Universe. 
They have been formed from the densest regions in the large-scale matter 
distribution of the Universe and have collapsed to form their own proper 
equilibrium structure. Their form can be well assessed by observations 
and well described by theoretical modelling \citep[e.g.][]{Sarazin88, 
Bahcall88, Voit05, Boehringer06, Ota12}.
X-ray and optical observations show that galaxy clusters are clearly defined 
connected structural entities, where the diffuse X-ray emission from the hot 
intracluster medium (ICM) trace the whole structure of the cluster. 
They are excellent giant laboratory sites for several astrophysical studies, 
for example, investigating galaxy evolution in their dense environments 
\citep[e.g.][]{Dressler80, Goto03}, evolution of the dynamical and thermal 
structure \citep[e.g.][]{Balestra07, Maughan08, Anderson09},  chemical 
enrichment of the intracluster medium \citep[e.g.][]{Cora06, Heath07}, 
studying lensed high-redshift background galaxies 
\citep[e.g.][]{Metcalfe03, Santos04, Bartelmann10}, and investigating the 
evolution of the Universe to test the cosmological models 
\citep[e.g.][]{Rosati02, Reiprich02, Voit05, Vikhlinin09a, Allen11}.  

Owing to the multi-component nature of galaxy clusters, they can be observed 
and identified through multiple observable signals across the electromagnetic 
spectrum. Tens of thousands of galaxy clusters have been identified by 
detecting their galaxies in the optical and near-infrared (NIR) bands 
\citep[e.g.][]{Abell58, Abell89, Zwicky61, Gladders05, Merchan05, Koester07, 
Wen09, Hao10, Szabo11, Geach11, Durret11, Wen12, Gettings12, Rykoff13}. Recently, 
several galaxy cluster surveys have been conducted at mm wavelengths using 
the Sunyaev-Zeldovich (SZ) effect based on observations made by several 
instruments, for example, the Atacama Cosmology Telescope \citep[ACT,][]
{Hasselfield13}, the South Pole Telescope \citep[SPT,][]{Reichardt13}, 
and the Planck Satellite \citep[][]{Planck13}. These surveys have provided 
cluster samples that contain several hundreds of SZ-selected clusters.     
 
X-ray cluster surveys provide pure and complete cluster catalogues, in addition,  
their X-ray observables correlate tightly with masses of clusters 
\citep[e.g.][]{Allen11}. Several hundreds of galaxy clusters were detected 
in X-rays based on previous X-ray missions mainly from ROSAT data 
\citep[e.g.][]{Ebeling98, Boehringer04, Reiprich02, Ebeling10, Rosati98, 
Burenin07}. The current X-ray telescopes (XMM-Newton, Chandra, Swift/X-ray) 
provide contiguous cluster surveys for small areas \citep[e.g.][]{Finoguenov07, 
Finoguenov10, Adami11,Suhada12}, in addition to serendipitous cluster surveys  
\citep[e.g.][]{Barkhouse06, Kolokotronis06, Lamer08, Fassbender11, Takey11, 
Mehrtens12, Clerc12, Tundo12, de-Hoon13, Takey13}. So far, these surveys have 
provided a substantial cluster sample of several hundreds up to a redshift of 1.57.   

We have conducted a systematic search for X-ray detected galaxy clusters based 
on XMM-Newton fields that are located in the footprint of the SDSS-DR7. The 
catalogue of serendipitously detected sources (extended) in XMM-Newton EPIC 
images was the basic database from which we selected a list of X-ray cluster 
candidates, comprising 1180 objects.
The main goal of the survey is to construct a large catalogue of newly 
discovered X-ray emitting groups and clusters. Due to the higher sensitivity 
of XMM-Newton the compiled cluster sample extends ROSAT cluster samples to 
fainter X-ray fluxes. The sample, which comprised galaxy groups and clusters,  
allows us to investigate the evolution of X-ray scaling relations as 
well as the correlation between the X-ray and optical properties. 
Other long-term goals of the survey are the selection of distant clusters 
beyond the SDSS detection limit and, in general terms, the preparation for 
the eROSITA mission, which will uncover a similar cluster population as 
in our survey.

The main way to obtain the cluster redshifts is based 
on the optical data. This can be achieved by either cross-matching the X-ray 
cluster candidates with the available optically selected galaxy cluster 
catalogues in the literature or by measuring the cluster photometric 
redshifts based on galaxy redshifts given in the SDSS catalogues. Using these 
two methods, we were able to establish an optically confirmed cluster sample 
comprising 530 groups/clusters with redshift measurements. From these  
optically confirmed groups/clusters with redshift measurements, we derived 
their X-ray luminosities and temperatures and investigated the X-ray 
luminosity-temperature relation. The selection criteria of the X-ray cluster 
candidates and redshift measurements as well as the X-ray properties of the 
optically confirmed sample were described in more detail by 
\citet[][Paper I, Paper II, hereafter]{Takey11, Takey13}.

In this work, we compile a sample of X-ray detected galaxy clusters among 
the X-ray cluster candidate list that are associated with luminous red 
galaxies (LRGs), which have spectroscopic redshift measurements out to 0.8 
in the SDSS-DR10. We present the procedure we used for constructing this 
cluster sample that is spectroscopically confirmed and for measuring their 
redshifts. We also present estimates of X-ray bolometric luminosity and 
luminosity-based mass at $R_{500}$ (the radius at which the cluster mean 
density is 500 times the critical density of the Universe at the cluster 
redshift) of the cluster sample.

The compiled cluster sample can be used to investigate various relations among 
the cluster physical properties, for example, the correlations between the 
properties of the BCG and its hosting cluster. By measuring the X-ray temperature 
of the cluster sample, one can extend the \ltr relation in Paper II to slightly  
higher redshifts. Moreover, the cluster sample is expected to permit studies of the 
relations between the cluster optical properties (richness and luminosity) 
and the cluster X-ray properties (X-ray temperature, luminosity, and mass). 
These correlations will be discussed in an upcoming paper.

The article is organized as follows: Section 2 gives a short overview on the 
selection procedure of the X-ray cluster candidates. In Section 3, we describe 
the construction of the cluster sample associated with LRGs and their redshift 
measurements. The X-ray parameters of the cluster sample are presented in 
Section 4. The summary of the paper is presented in Section 5. Throughout 
this paper, we used the cosmological parameters  $\Omega_{\rm M}=0.3$, 
$\Omega_{\Lambda}=0.7$ and $H_0=70$\ km\ s$^{-1}$\ Mpc$^{-1}$.


\section{Description of the X-ray cluster candidates}

Galaxy clusters are simply identified among the X-ray sources as X-ray 
luminous, spatially extended, extragalactic sources \citep{Allen11}. 
The largest X-ray source catalogue so far is the XMM-Newton serendipitous 
source catalogue, which has been created by the XMM-Newton Survey Science 
Centre (SSC). The latest edition of this catalogue is the 
3XMM-DR4\footnote{\url{http://xmmssc-www.star.le.ac.uk/Catalogue/xcat_public_3XMM-DR4.html}},   
which contains 372728 unique X-ray sources drawn from 7427 
XMM-Newton EPIC observations made between 2000 February 3 and 2012 December 8. 
Our survey was based on the previous edition of the XMM-Newton serendipitous 
source catalogue 
\cite[2XMMi-DR3\footnote{\url{http://xmmssc-www.star.le.ac.uk/Catalogue/xcat_public_2XMMi-DR3.html}},]
[]{Watson09} that was compiled based on 4953 XMM-Newton observations made between 
2000 February 3 and 2009 October 8. The 2XMMi-DR3 catalogue comprises 353191 
detections corresponding to 262902 unique sources. Of these detections, 
30470 are extended detections, which include both real and spurious 
extended sources as well as multiple detections of the same sources.

We selected the X-ray cluster candidates from reliable extended sources 
(with no warning about being spurious) in the 2XMMi-DR3 catalogue at high 
galactic latitudes, $|b| > 20\degr$. 
The survey was constrained to the XMM-Newton fields that were located in the 
footprint of the SDSS-DR7 to be able to measure the optical redshifts 
of the possible optical counterparts. The overlap area of XMM-Newton fields and 
the imaging area of the SDSS-DR7 is 210 deg$^2$. After excluding possible 
spurious X-ray extended detections and low-redshift galaxies that appear 
resolved at X-ray wavelengths through visual inspections of the X-ray images 
and the X-ray-optical overlays, the X-ray cluster candidate list comprised 
1180 objects. The selection procedure was described in more detail in papers 
I and II.         

\begin{figure}
  \resizebox{\hsize}{!}{\includegraphics{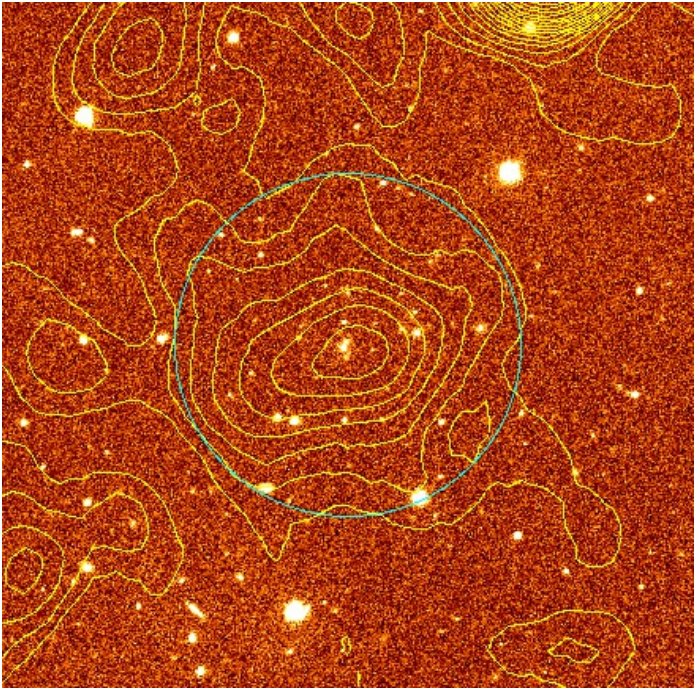}}
  \caption{SDSS image of the cluster 2XMMi J143742.9+340810 at a redshift 
of 0.5446, with X-ray surface brightness contours (0.2 - 4.5 keV) overlaid 
in yellow. The plotted cyan circle has a radius of one arcmin around the 
X-ray position. The field of view is $4'\times 4'$ centred on the X-ray 
emission peak.} 
  \label{f:275341_overlay}
\end{figure}

In this paper, we identify a subsample of these X-ray 
cluster candidates associated with LRGs that have spectroscopic redshifts 
in the SDSS-DR10 to construct a sample with spectroscopic confirmations. 
As an example, Fig.~\ref{f:275341_overlay} shows a galaxy cluster, 
2XMMi J143742.9+340810, at a redshift of 0.5446, associated with two LRGs as 
cluster member galaxies with available spectroscopic redshifts. We use this 
cluster to show the procedure of measuring the redshift in the next section.

\section{Clusters associated with spectroscopically targeted LRGs in SDSS-DR10}

Generally, the brightest cluster galaxies (BCGs) are elliptical massive 
galaxies and reside near the cluster centre of mass. The BCGs tend to be 
very luminous and red galaxies \citep[e.g.][]{Postman95, Eisenstein01, 
Wen12}, therefore they have a good chance to be selected for BOSS 
spectroscopy in SDSS-III as members of the LRG sample, which in turn allows 
a straightforward spectroscopic confirmation of associated galaxy clusters 
\citep[e.g.][]{Goto02, Mehrtens12}.

We identified the LRGs from the SDSS-DR10 data that are cluster member galaxies of 
the X-ray cluster candidates. The spectroscopic redshifts of these cluster members 
were used to measure the cluster redshift. In the next subsections, we describe 
how we selected the LRGs from the recent data of the SDSS and the procedure we 
followed to measure the cluster redshift. We also present some statistical 
properties of the constructed cluster sample, such as the redshift distribution 
and the projected offsets of the BCGs from the X-ray positions. Then 
a comparison of the current redshift measurements with the published ones 
is presented. Finally, we give a general overview of the total optically 
confirmed cluster sample from our survey. 


\subsection{Luminous red galaxy sample}
The Sloan Digital Sky Survey (SDSS) has been in continuous operation since 2000. 
The latest data release so far from the SDSS is Data Release 10 
\citep[SDSS-DR10,][]{Ahn13}, which provides the spectroscopic data from the 
SDSS-III's Baryon Oscillation Spectroscopic Survey (BOSS) as well as imaging 
and spectroscopic data from the previous SDSS data releases. SDSS-DR10 provides 
optical spectra for 1880584 galaxies. BOSS is an ongoing project and its  
current data release includes 927844 galaxy spectra, in addition to thousands 
of quasar and stellar spectra, which were selected over 6373.2 deg$^2$. One aim 
of BOSS is to  obtain spectra of 1.5 million galaxies in the redshift range of 
$0.15 < z < 0.8$  distributed over 10000 deg$^2$. Therefore, it is a valuable 
resource to obtain spectroscopic redshifts for luminous cluster galaxies.

Luminous red galaxies are among the most luminous red galaxies that can be 
observed up to redshift 0.8 with SDSS equipment.  
As a starting point for identifying LRGs that are expected 
to be cluster galaxies, we selected all galaxies within 10 armin of the X-ray 
position, which is the maximum angular scale at the low-redshift end 
(500\,kpc at $z=0.04$) to be considered for our cluster candidate sample. 
In a second iteration step, we refined this search radius according to the 
initial cluster redshift estimate based on the redshift of the BCG candidate.       
The galaxies were selected from the {\tt galaxy} view table in the SDSS-DR10, 
which contains the photometric parameters measured for resolved primary objects, 
classified as galaxies.  

The photometric redshifts ($z_{\rm p}$) and, 
where available, the spectroscopic redshifts ($z_{\rm s}$) of the galaxy sample 
were also selected from the {\tt Photoz} and {\tt SpecObj} tables, respectively. 
The {\tt SpecObj} table includes spectroscopic redshifts that were measured 
from clean galaxy spectra taken by the new and old spectrographs in the SDSS 
projects. The extracted parameters of the galaxy sample included the coordinates, 
the (model and composite model) magnitudes in $r-$ and $i-$bands, the photometric 
redshifts, and, where available, the spectroscopic redshifts.
We used the magnitudes in the {\tt galaxy} table that are corrected for 
Galactic extinction following \citet{Schlegel98}. To clean the galaxy sample 
from faint objects beyond the completeness limits of SDSS or from galaxies with 
large uncertainty in $z_{\rm p}$, we only considered galaxies that have a model 
magnitude of $m_{r} \leq 22.2$ mag and $\bigtriangleup m_{r} < 0.5 $ mag and 
a relative error of photometric redshift of $\bigtriangleup z_{p}/z_{p} < 0.5$.  

The BOSS data include two main target galaxy samples; first the BOSS LOWZ 
galaxy sample with $z \le 0.4$; second the BOSS constant-mass CMASS  
galaxy sample with $0.4 < z < 0.8$. The target selection algorithms for galaxies 
in BOSS are significantly different from those used in the previous SDSS 
projects because of the different scientific goals \citep{Ahn12}. BOSS galaxy 
targets are significantly fainter than those in the previous SDSS 
projects with the aim of measuring large-scale clustering of galaxies at 
higher redshifts. 

To select a homogeneous luminous red galaxy sample from 
BOSS and previous SDSS data releases, we applied the same selection criteria 
on both data. We selected LRGs with available spectroscopic redshifts from the 
constructed galaxy sample within 10 arcmin from the X-ray positions.  
The applied selection criteria of LRGs were based on the colour and 
magnitude cuts that are described by Padmanabhan et al. (2014, in preparation) 
and given on the SDSS   
website\footnote{\url{http://www.sdss3.org/dr10/algorithms/boss_galaxy_ts.php}} 
as well as in Appendix A.
We also ensured that the selected objects are confirmed galaxies using 
the spectroscopic class parameter given in the {\tt SpecObj} table to 
exclude objects targeted as galaxies that were stars or 
quasars. The selected LRG sample was used to identify the BCGs of the X-ray 
cluster candidates as described in the following subsection.


\subsection{Optical identifications and redshift measurements}

\begin{figure}
  \resizebox{\hsize}{!}{\includegraphics{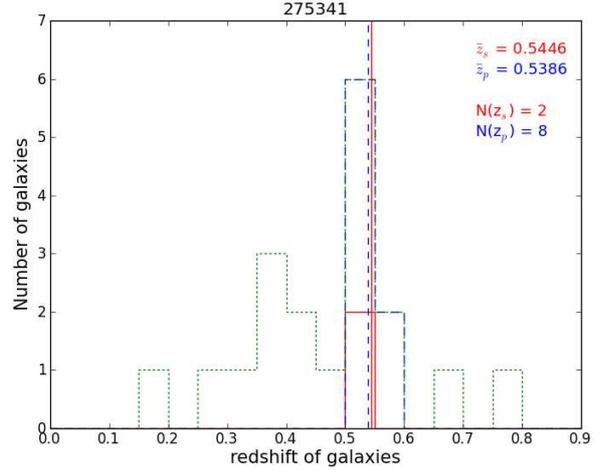}}
  \caption{Histogram of the spectroscopic ($z_{\rm s}$, red solid) 
and photometric ($z_{\rm p}$, blue dashed) redshifts of the cluster member 
candidates ($N_{z_{\rm s}}$ and $N_{z_{\rm p}}$) within 500 kpc of the example 
cluster, 2XMMi J143742.9+340810. The green dotted histogram represents the 
distribution of $z_{\rm p}$ of field galaxies within the same region. 
The cluster spectroscopic $\bar z_{\rm s}$ and photometric $\bar z_{\rm p}$ 
redshifts are represented by red solid and blue dashed vertical lines, 
respectively, and are listed in the upper corner.} 
  \label{f:275341_hist_DR10}
\end{figure}

\begin{figure}
  \resizebox{\hsize}{!}{\includegraphics{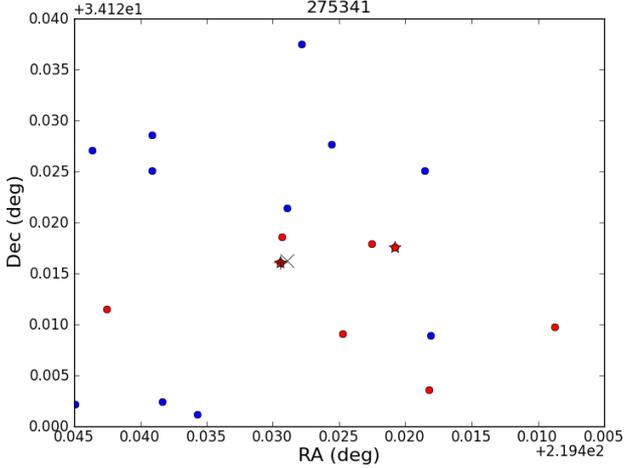}}
  \caption{Distribution on the sky of the cluster member candidates 
(red dots) and field galaxies (blue dots) within 500 kpc ($\sim$ 1.3 arcmin) 
from the X-ray position (marked by a black cross) of the example cluster, 
2XMMi J143742.9+340810. Note the different scale in 
Fig.~\ref{f:275341_SDSS}. The cluster galaxies with available $z_{\rm s}$ 
are marked by stars. The BCG candidate is marked by a black plus that  
has a projected separation from the X-ray position of $\sim$ 11 kpc. We only 
present galaxies with $ m_{r} \leq 22.2$ mag,  
$\bigtriangleup m_{r} < 0.5$ mag, and $\bigtriangleup z_{p}/z_{p} < 0.5$. } 
  \label{f:275341_dist_DR10}
\end{figure}

\begin{figure}
  \resizebox{\hsize}{!}{\includegraphics{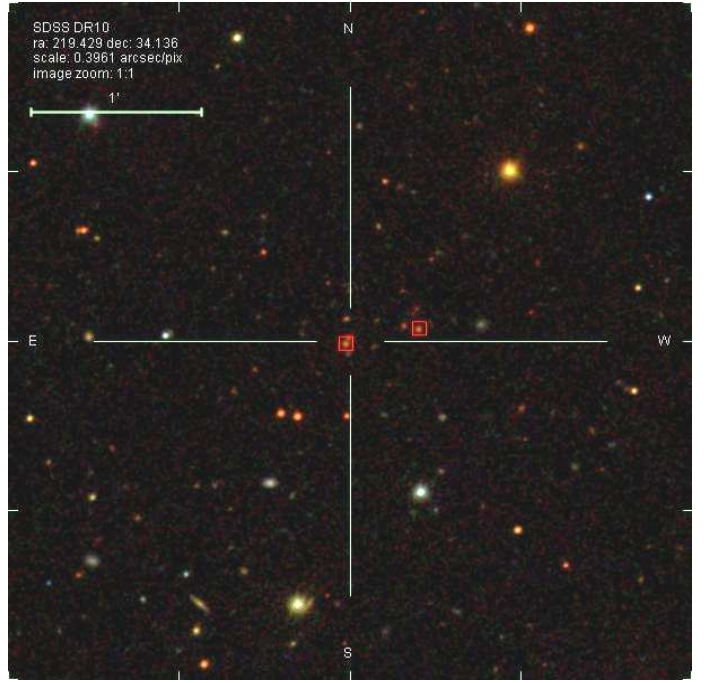}}
  \caption{SDSS colour image of the cluster 2XMMi J143742.9+340810  
with 4 arcmin a side centred on the X-ray position, which is marked by the 
cross-hair. Two cluster galaxies (LRGs) with spectra are marked by red 
squares. The measured spectroscopic redshift for this system is 0.5446, 
which is computed as the weighted average, $\bar z_{\rm s}$, of the 
spectroscopic redshifts of the two LRGs. } 
  \label{f:275341_SDSS}
\end{figure}

To measure the redshifts of cluster candidates, we firstly identified the 
BCG candidates, then we selected cluster member candidates with available 
similar $z_{\rm s}$ of the BCG's spectroscopic redshift. The procedure works  
as follows: 

\begin{enumerate}

\item We identified a BCG candidate as an LRG within 200 kpc (computed based 
on $z_{\rm s}$ of the LRG) from the X-ray position of the cluster candidate. 
If there was only one LRG, we considered it as the BCG candidate. If there 
were several LRGs, we divided them into groups with similar redshifts that  
are within a redshift interval of $\bigtriangleup z_{\rm s} = 0.01 $. When 
there was only one group, we chose the brightest LRG as the BCG candidate. If 
there was more than one group, the priority was given to the group with 
more members and the BCG candidate was selected as the brightest galaxy from 
this group. If the groups have the same count of LRGs, we initially selected the 
brightest galaxy in each group. Then the next-brightest LRG to the X-ray 
emission peak was regarded as the BCG candidate. The maximum number of LRGs 
within 200 kpc from the X-ray centre was four. 

At low redshifts, the search radius of 200 kpc subtends a large 
angle on the sky and might cause an incorrect association of 
LRGs with the X-ray cluster candidates. Therefore, we set a maximum angular 
separation limit on the BCGs offset from the X-ray emission peak of 90 arcsec.
The search radius of 200 kpc was used since we found that 90 percent of the 
BCGs in Paper II had an offset lower than 200 kpc. The search radius 
(200 kpc) used here is slightly larger than the search radius (175 kpc) 
used by \citet{Mehrtens12} to identify LRGs that are part of the cluster.
These authors assigned the spectroscopic redshift of an LRG or a group of 
LRGs within 175 kpc from the X-ray position as the cluster redshift.

\item We identified the cluster member candidates (not necessarily LRGs) 
within 500 kpc from the X-ray peak based on the spectroscopic redshift of the 
identified BCG candidate, $z_{\rm s,\rm BCG}$. 
The cluster galaxies with available $z_{\rm s}$ were selected within a small 
redshift interval of $z_{\rm s,\rm BCG} \pm 0.01$. While the cluster member 
candidates with only $z_{\rm p}$ were selected within a slightly larger
redshift interval of $z_{\rm s,\rm BCG} \pm 0.04(1+z_{\rm s,\rm BCG})$.     
The distribution of the redshifts of the cluster member candidates and field 
galaxies for the example cluster is shown in Fig.~\ref{f:275341_hist_DR10}. 
The redshift interval we used to identify the cluster members with $z_{\rm p}$ 
gives 80 percent of the cluster members \citep{Wen09}. These authors also showed 
that a radius of 500 kpc gives a high overdensity level and a low false-detection 
rate. The identified BCG candidate could be a cluster galaxy fainter 
than the first BCG, thus we re-identified the likely BCG as the brightest 
galaxy among the cluster member candidates within 500 kpc.

\item We computed the spectroscopic, $\bar z_{\rm s}$, and photometric, 
$\bar z_{\rm p}$, redshift of a cluster as a weighted average of the 
spectroscopic and photometric redshifts of the cluster member candidates 
within 500 kpc, respectively. 
The weights are given as $w_i = 1/(\bigtriangleup z_{s,\,i})^2$ in computing
$\bar z_{\rm s}$ and as $w_i = 1/(\bigtriangleup z_{p,\,i})^2$ in determining
 $\bar z_{\rm p}$. 
If there was only one cluster galaxy (LRG) with available spectroscopic 
redshift $z_{\rm s}$, we considered its $z_{\rm s}$ as the cluster redshift.    

\item We accepted the optical counterpart and the redshift measurement of an 
X-ray cluster candidate if the optical detection passed the quality assessment 
performed through a visual inspection process. 
We compared the 
identified BCG and cluster member candidates with the corresponding SDSS 
colour image of the same field. The sky distribution of cluster members 
of the example cluster is shown in Fig.~\ref{f:275341_dist_DR10}, 
while Fig.~\ref{f:275341_SDSS} shows the corresponding SDSS colour image. 
From both images, it was obvious that the algorithm picked the correct  
associated LRGs (and thus the BCG too) and the luminous cluster member 
candidates. The fainter cluster galaxies were not considered in 
Fig.~\ref{f:275341_dist_DR10} because of the magnitude limit we used to 
create the galaxy sample or they were not detected at all in 
Fig.~\ref{f:275341_SDSS} because of the detection limit of the 
SDSS imaging.

\end{enumerate}

The current procedure yielded an initial list of optical counterparts that 
comprised 415 candidates. 
We regard about 8 percent of the initial candidates as doubtful for 
two main reasons, overlapping clusters and unrelated nearby galaxies. We 
found cases of most likely overlapping clusters where the algorithm probably 
picked an incorrect cluster since an LRG of this not-so-distant cluster happened 
to lie closer to the X-ray position. Another reason for an incorrect redshift 
estimate is the identification of 
a bright foreground galaxy as BCG candidate for a distant cluster candidate, 
where no detected cluster galaxies are found around the X-ray emission 
peak. We excluded such systems from the initial cluster candidate sample. 
Therefore, the estimated misidentification fraction of the resulting 
optical counterpart list using the current cluster identification procedure 
is about 8 percent. These systems were removed from the sample without 
additional attempts to correct for or quantify their effect. 
The final list of the optically validated cluster sample included 383 
systems with spectroscopic confirmation based on at least one 
spectrum of an LRG.

\subsection{Optically validated cluster sample}

\begin{figure}
  \resizebox{\hsize}{!}{\includegraphics{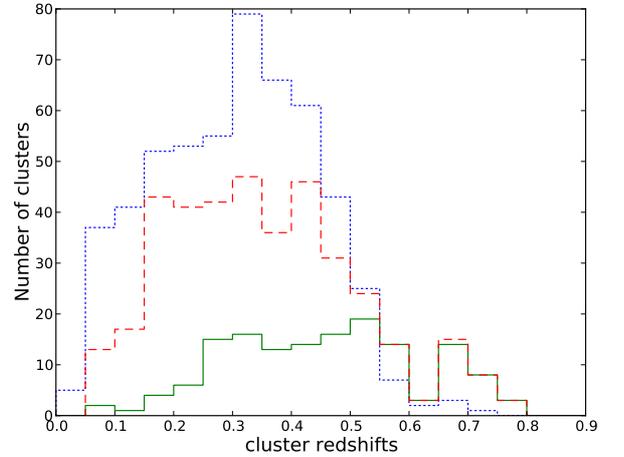}}
  \caption{Distribution of measured spectroscopic redshifts of the 
cluster sample associated with LRGs that have spectra is presented by the 
red dashed line, and a subsample of it based only on BOSS data is indicated 
by the green solid line, while the redshift distribution of the optically 
confirmed cluster sample in Paper II is presented by the blue dotted line.} 
  \label{f:Hist_zc_DR10}
\end{figure}

The redshifts of the optically confirmed cluster sample (383 systems) span 
a wide range from 0.05 to 0.79 with a median of 0.34. Among this sample, 147 
clusters are spectroscopically confirmed based only on the SDSS-III BOSS data. 
The redshift distribution of the current cluster sample (383 systems), the 
subsample based only on BOSS survey (147 objects), and the optically confirmed 
cluster sample (530 systems) in Paper II are shown in Fig.~\ref{f:Hist_zc_DR10}. 
The redshift distribution of the current cluster sample is dependent on the 
selection of LRG targets in SDSS projects and the selection of X-ray 
cluster candidates from the 2XMMi-DR3 catalogue as well as the cluster 
identification procedure we used here to construct the cluster sample.
The drop of the redshift histogram at $z=0.60-0.65$ might be a result of 
a combination of these effects or possibly the low-statistics regime of the 
survey.

The objects in common between the two samples (Paper II sample and the current 
one) are 316 systems, see the next subsection for the redshift comparison. 
The current cluster sample includes 40 more distant clusters beyond 
$z = 0.5$ than the distant sample in Paper II. As shown in 
Fig.~\ref{f:Hist_zc_DR10}, these distant clusters were detected based 
on BOSS data, thanks to the data release of BOSS in the SDSS-DR10. 
Additionally, the current cluster sample extends the confirmed cluster 
sample in Paper II by 67 systems, of these 52 systems are newly discovered  
galaxy clusters. 
Therefore, the current cluster sample is complementary to that 
presented in Paper II in terms of size and redshift range. Both samples can be 
combined for future investigations of X-ray scaling relations as well 
as X-ray-to-optical relations.

\begin{figure}
  \resizebox{\hsize}{!}{\includegraphics{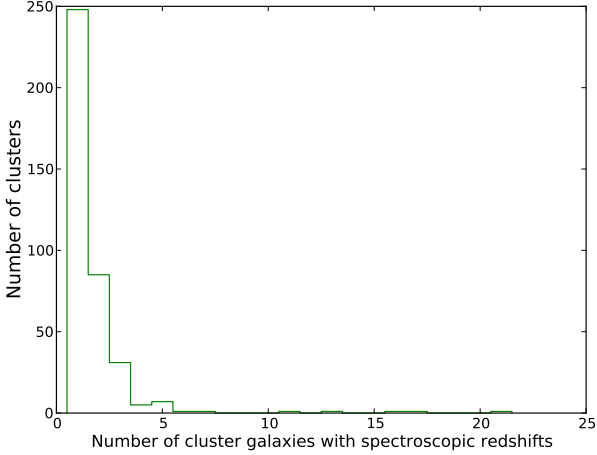}}
  \caption{Distribution of the number of identified spectroscopic cluster 
galaxies per system within 500 kpc from the X-ray positions for the optically 
validated cluster sample. The bin size of the histogram is one. }
  \label{f:Hist_Nzs_DR10}
\end{figure}

The majority of the clusters in the sample have one or two cluster member 
galaxies with $z_{\rm s}$, while few clusters have three member galaxies or 
more with available $z_{\rm s}$. The distribution of the cluster galaxies 
with $z_{\rm s}$ per cluster of the cluster sample is shown in 
Fig.~\ref{f:Hist_Nzs_DR10}. 
The cluster member galaxies are not complete for the distant systems 
in our cluster sample. Faint cluster galaxies were not considered or detected 
because of the magnitude limit (SDSS completeness limit) used in the current 
procedure or the limited data depth of the SDSS imaging, respectively.  
For instance, only the BCG (LRG) was identified for the most distant 
cluster in the sample at a redshift of 0.79. 

\begin{figure}
  \resizebox{\hsize}{!}{\includegraphics{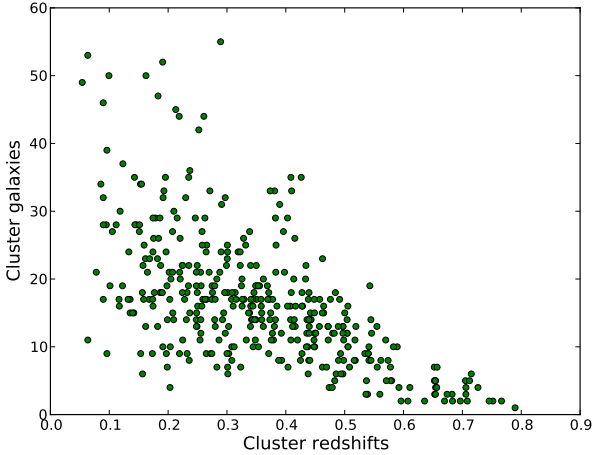}}
  \caption{Cluster member galaxies plotted against the cluster spectroscopic 
redshifts. These cluster member candidates were selected based on 
their $z_{p}$ within 500 kpc from the X-ray emission peak } 
  \label{f:Nzp_zs_DR10}
\end{figure}

Figure~\ref{f:Nzp_zs_DR10} shows the scatter plot between the count 
of identified cluster member candidates per system with $z_{\rm p}$ and 
within 500 kpc from the X-ray position and the cluster redshift. 
It is shown that distant clusters have 
only a few identified luminous cluster galaxies, mainly because of the sensitivity 
limits of the SDSS. We kept these objects in our cluster candidate list, even 
with only two member galaxies, for the summed evidence of extended X-ray emission 
positioning that coincides with an LRG with spectra. To determine a complete cluster 
richness, one needs to follow-up these systems to obtain deep imaging data.

\begin{figure}
  \resizebox{\hsize}{!}{\includegraphics{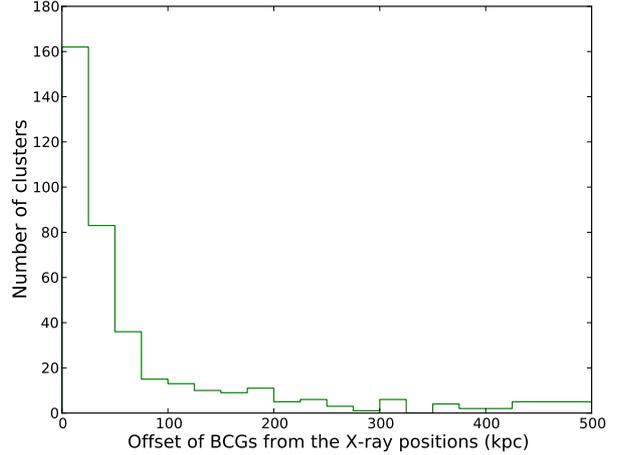}}
  \caption{Histogram of the projected separations between the probable BCGs 
and the X-ray emission peaks of the cluster sample.} 
  \label{f:Hist_offset_DR10}
\end{figure}

Based on the cluster redshift and the angular separation of the BCGs to the 
X-ray peaks, we computed their projected offsets. The distribution of the 
projected separations between the probable BCGs and the X-ray emission 
peaks is shown in 
Fig.~\ref{f:Hist_offset_DR10}. We found that the majority of the BCGs (about  
90 percent) have offsets smaller than 200 kpc, which agree with the offsets 
of BCGs sample in Paper II. By using the current selection procedure of the 
BCG as a cluster member galaxy within 500 kpc from the X-ray centers, the 
maximum offset is about 500 kpc. The large offset of the BCGs from the X-ray 
centroids might appear in systems with an ongoing merger or in dynamically 
active clusters \citep{Rykoff08}.

The distribution of positional offsets of the BCGs from the X-ray positions 
might be biased by the initial LRG (BCG candidate) selection, which was 
required to lie within a projected distance of 200 kpc.
Since we excluded apparent doubtful systems from the initial cluster sample, 
the distribution is less affected by identifying incorrectly associated BCGs.  
The median offset of the BCG for the cluster sample is 33 kpc. We  
investigated a possible evolution of the BCG offset with the cluster 
redshift. The cluster sample was divided into three subsamples with redshift 
bins of $0.05 \le z < 0.30, 0.30 \le z < 0.55$, and $0.55 \le z < 0.80$ and 
number of clusters 156, 184, and 43, respectively. The median BCG offsets for 
the low, intermediate, and distant redshift subsamples were 23 kpc, 42 kpc, 
and 45, respectively. These numbers indicate a trend of increasing 
positional offset of the BCG with increasing redshift.

We compared the BCGs properties of the cluster sample (383 systems) with the 
selection criteria of the spectroscopically targeted LRGs in BOSS survey, 
see Appendix A. We found that 92 percent of the BCGs fulfil the colour and 
magnitude cuts of the LRGs in BOSS.
This high percentage is affected by our strategy of the cluster identification, 
which selected a cluster galaxy as an LRG within 200 kpc from the X-ray emission peak.

\subsection{Comparison with published redshifts}

The largest optically selected galaxy cluster sample so far was compiled 
by \citet[][WHL12 hereafter]{Wen12}, based on overdensities of galaxies in 
photometric redshift space from the SDSS-DR8 data. It comprises 132684 
clusters with photometric redshift measurements in the range of 
$0.05 \le z < 0.8$. The catalogue also provides spectroscopic redshifts 
for about 30 percent of the BCGs. Cross-matching our sample with the 
WHL12 catalogue yielded 188 common clusters, of these, 131 systems have 
spectra for the BCGs in their catalogue. 
We also queried the NASA Extragalactic Database (NED) for available redshift 
measurements for the remainder of the cluster sample. As a result, 76 clusters
with redshift estimates from different projects were found. In total, 264 
clusters are previously known in the literature mostly as optically selected  
galaxy clusters.

\begin{figure}
  \resizebox{\hsize}{!}{\includegraphics{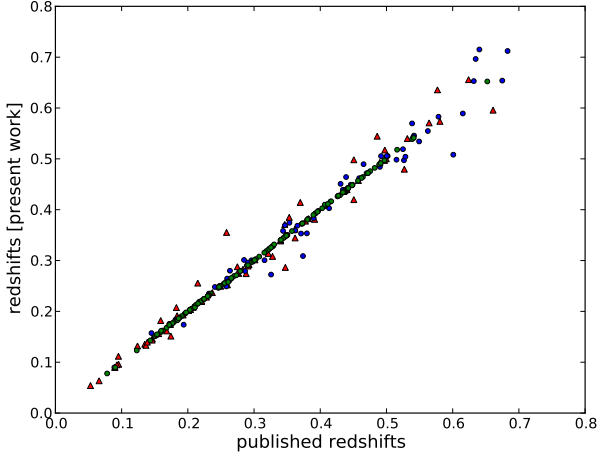}}
  \caption{Comparison of the present spectroscopic redshift measurements with 
the published BCG spectroscopic (green dots) and cluster photometric (blue dots) 
redshifts by \citet{Wen12} and the available (either spectroscopic or 
photometric) redshifts from the NED (red triangle).} 
  \label{f:zpre_zpub}
\end{figure}

Figure~\ref{f:zpre_zpub} shows the comparison between the present redshift 
measurements and the WHL12 ones as well as the available redshifts from the NED.
The good agreement between the current redshift measurements and the published 
values is clear. The differences between the two measurements, 
$\bigtriangleup z = z_{\rm present} - z_{\rm published}$, have a mean and 
standard deviation of 0.0009 and 0.0170, respectively.  

We also compared the current redshift measurements with the published ones of 
the optically confirmed cluster sample from our ongoing survey (Paper II). 
There are 316 clusters in common between the two samples, of these, 238 have 
spectroscopic redshifts for at least one cluster galaxy in Paper II.
The current procedure spectroscopically confirmed the remainder 
of the common sample with only photometric redshifts (78 systems). 
We noted that the current procedure did not identify the whole sample 
in Paper II with spectroscopic confirmations (310 clusters). This is because 
we used the criterion of having an LRG with $z_{\rm s}$ within 200 kpc. 
In addition, in Paper II we used the spectroscopic data from the SDSSI/II 
projects, which provides a galaxy sample with $z_{\rm s}$ including LRGs 
as well as a magnitude-limited galaxy sample that are not necessarily LRGs.      

\begin{figure}
  \resizebox{\hsize}{!}{\includegraphics{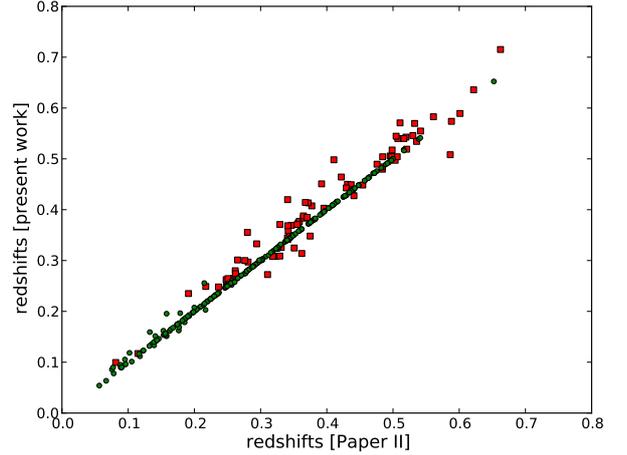}}
  \caption{ Comparison of the present spectroscopic redshift measurements with 
the measured spectroscopic redshifts (green dots) and photometric ones 
(red squares) from Paper II.} 
  \label{f:zp3_zp2}
\end{figure}

Figure~\ref{f:zp3_zp2} shows the comparison of the present redshift 
measurements with the values from Paper II of the common sample. The two 
measurements agree well. The mean and standard 
deviation of the differences between the measured values, 
$\bigtriangleup z = z_{\rm present} - z_{\rm Paper\,II}$, are 0.0039 and 0.0150, 
respectively. Only 4 percent of the common sample have redshift 
differences of $2\sigma < |\bigtriangleup z| \le 4\sigma$, where 
$\sigma = 0.02$ is the uncertainty of the measured photometric redshifts 
in Paper II. Systems with such redshift differences had only 
photometric redshifts in our previous work.  
 
We also noted that the present spectroscopic redshifts are not identical with 
the spectroscopic redshifts in Paper II for a few cases.  There are eleven  
clusters with  redshift differences of $0.01 < |\bigtriangleup z| < 0.04$.
This is because cluster galaxies in Paper II were selected within a redshift 
interval of $z_{\rm p,\rm BCG} \pm 0.04(1+z_{\rm p,\rm BCG})$, then the cluster 
spectroscopic redshift was measured as the weighted average of the available 
spectroscopic redshifts of these identified cluster galaxies. This led to 
selecting galaxies with available $z_{\rm s}$ that have redshifts outside 
the redshift interval we used in this work ($z_{\rm s,\rm BCG} \pm 0.01$, 
see subsection 3.2).

\subsection{Combined optically confirmed cluster sample from the 2XMMi/SDSS 
cluster survey}

Based on the photometric redshifts of galaxies in the SDSS-DR8, we have 
optically identified the counterparts of 530 galaxy groups/clusters in 
Paper II including the cluster sample in Paper I . In the present work, 
we constructed a cluster sample of 383 systems based on the spectroscopic 
redshifts of galaxies in the SDSS-DR10. 
The spectroscopically confirmed cluster sample extends the optically confirmed 
cluster sample in Paper II by 67 objects. Therefore, out of 1180 X-ray cluster 
candidates we optically confirmed 597 galaxy groups/clusters 
(51 percent) and measured their redshifts. 

This is the largest X-ray selected galaxy cluster catalogue so far based on 
the archival XMM-Newton observations. The XMM-Newton sensitivity and the 
available deep XMM-Newton fields allowed us to detect clusters down to X-ray 
flux of $\sim 10^{-15}$\ erg\ cm$^{-2}$\ s$^{-1}$ in [0.5 - 2.0] keV. However,  
this is not a flux-limited cluster sample since we included XMM-Newton 
observations with different exposure times in our survey.    
The redshifts of the full cluster sample span a wide range from 0.03 to 
0.79 with a median of 0.34. Concerning spectroscopic redshifts, out of 597 
confirmed clusters 455 systems (76 percent) have a spectroscopic confirmation 
based on at least one member galaxy with a spectrum.

Overall, we are left with 583 unconfirmed X-ray cluster candidates (49 percent). 
These remaining candidates are either distant clusters beyond the SDSS detection 
limit, systems with low richness that are groups with few members, or possibly 
spurious detections.



\section{X-ray parameters}

In Paper II, we provided two subsamples of clusters: (i) a cluster subsample 
comprising 345 objects with their X-ray spectroscopic temperature and 
flux from the spectral fitting, and (ii) a cluster subsample consisting of 
185 systems with their X-ray flux obtained from the 2XMMi-DR3 catalogue, 
because their X-ray data are insufficient for spectral fitting. For both 
subsamples, we also estimated the X-ray bolometric luminosity $L_{500}$ 
using an iterative method based on scaling relations. This iterative  
procedure was the same for the two subsamples but with different inputs,  
see Section 4 in Paper II for more detail about this procedure. We found 
a good agreement between the derived $L_{500}$ for the clusters in common 
between the two subsamples.

In the current work, we estimate the X-ray parameters ($L_{500}$ and $M_{500}$) 
for the present optically validated cluster sample (383 systems) based on the 
integrated $\beta$ model fluxes that were given in the 2XMMi-DR3 catalogue. 
Firstly, we compute the X-ray luminosities using the catalogue fluxes and 
the measured redshifts. Secondly, we convert the computed luminosity to 
bolometric luminosities $L_{500}$. Finally, the estimated $L_{500}$ 
are used to compute the cluster masses $M_{500}$.


\subsection{X-ray luminosity and mass}

For each system in the current cluster sample, we obtained its X-ray   
aperture-corrected, $\beta$-model, flux that was calculated by the SAS tasks 
{\tt emldetect} from the 2XMMi-DR3 catalogue \citep{Watson09}. Here we used 
the combined EPIC (MOS1, MOS2, PN) flux in (0.5-2.0 keV), $F_{\rm cat,\,0.5-2}$, 
and its propagated error. Then we computed the X-ray luminosity in 
(0.5-2.0 keV), $L_{\rm cat,\,0.5-2}$.

\begin{figure}
  \resizebox{\hsize}{!}{\includegraphics{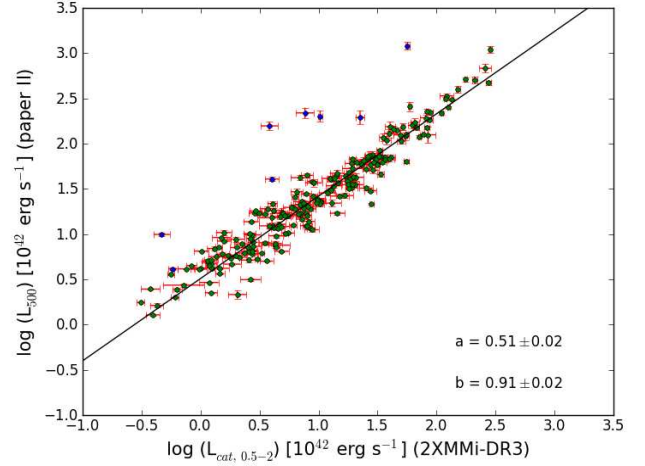}}
  \caption{X-ray bolometric luminosity $L_{500}$ from Paper II plotted 
against the present luminosity $L_{\rm cat,\,0.5-2}$ of 218 clusters in 
common between the current cluster sample and the sample in Paper II with 
reliable X-ray parameters. The solid line represents the best fit derived 
using the BCES orthogonal regression method after excluding eight outliers 
that are represented by blue dots. The intercept (a) and the slope (b) of 
the linear relation are written in the lower left corner.} 
  \label{f:L500-Lx}
\end{figure}

To convert $L_{\rm cat,\,0.5-2}$ into bolometric luminosity $L_{500}$, we used 
the properties of the cluster sample with reliable X-ray spectroscopic
parameters in Paper II. Among the current optically validated cluster sample, 
there are 218 clusters with X-ray spectroscopic parameters from the spectral 
fitting were already published in Paper II. We compared bolometric $L_{500}$ 
from Paper II and $L_{\rm cat,\,0.5-2}$ from the current procedure of these 
common systems, as shown in Fig.~\ref{f:L500-Lx}. It shows a linear relation 
between the two luminosity measurements with a few outliers (about 4 percent), 
which are contaminated by point sources. The best-fit linear relation after 
excluding these outliers derived using the BCES orthogonal regression method 
\citep{Akritas96} is represented by the solid line in Fig.~\ref{f:L500-Lx} 
and is given by 
\begin{equation}
  \log\ (L_{500}) = (0.51 \pm 0.02) + (0.91 \pm 0.02)\ \log\ (L_{\rm cat,\,0.5-2}).  
\end{equation}
The error in $L_{500}$ was calculated by including the error measurement of  
$L_{\rm cat,\,0.5-2}$, the intrinsic scatter in this relation (Eq.~1), and 
the propagated errors caused by the uncertainty in the slope and the intercept.
The intrinsic scatter value of the relation (Eq.~1) is $0.15 \pm 0.01$,
which was estimated following  the method used by \citet{Pratt09}.

Eq. 1 provides a quick method to derive bolometric $L_{500}$ based on the measured 
$L_{\rm cat,\,0.5-2}$. This scaling relation implicitly includes the bolometric
correction (with scatter introduced by the X-ray temperature and luminosity 
in the energy band [0.5-2.0] keV) as well as aperture flux extrapolation to 
$R_{500}$ (with scatter introduced by different surface brightness profile 
parameters).

The estimated bolometric $L_{500}$ was used to compute the X-ray 
luminosity-based mass using the $L_{500}-M_{500}$ relation published by 
\citet{Pratt09} of the form  
\begin{equation}
 M_{500} = 2 \times 10^{14} M_{\odot}\ \bigl(
 \frac{h(z)^{-7/3}\ L_{500}}{1.38\times10^{44}\ \rm erg\ \rm s^{-1}} \Bigr)^{1/2.08},  
\end{equation}
where $h(z)$  is the Hubble constant normalised to its present-day value, 
$h(z) = \sqrt{\Omega_{\rm M} (1+z)^{3} + \Omega_{\Lambda}}$. 
We calculated the error in $M_{500}$ that has contribution from the error in 
$L_{500}$, the intrinsic scatter in Eq.~2, and the propagated errors for 
the uncertainty in relation's slope and intercept. 
Finally, $M_{500}$ was used to compute $R_{500}$ as
\begin{equation}
 R_{500} = \sqrt[3]{3 M_{500} / 4\pi 500 \rho_{c}(z)},
\end{equation} 
where $\rho_{c}(z)$ is the critical density, 
$\rho_{c}(z) = h(z)^{2} 3 H_0^{2} / 8 \pi G$ .

For the whole current cluster sample (383 systems), we derived  
$L_{\rm cat,\,0.5-2}$, $R_{500}$, $L_{500}$ and $M_{500}$, as described above. 
Among this sample, there are 316 clusters in common with the whole sample 
(530 objects) in Paper II. The current sample extended the optically confirmed 
cluster sample from our survey by 67 objects and added several clusters 
at higher redshift up to $z = 0.79$.  Since we provided the measurements of 
$L_{500}$ and $M_{500}$ for the common sample in Paper II, we only present 
the catalogue of the extended cluster sample (67 systems) listing their 
optical and X-ray parameters. In addition to this new sample, we also provide 
the spectroscopic redshifts and the X-ray properties for 78 clusters among 
the common sample in Paper II that previously had only photometric redshifts. 
These two subsamples are compiled in Table~\ref{tbl:catalog_sample} 
with a note referring to each subsample.

Table~\ref{tbl:catalog_sample}, available in full form  at the CDS, lists  
the new cluster sample (67 objects) from the current 
work in addition to a subsample (78 systems) that has only photometric 
redshifts in Paper II. The X-ray parameters are estimated based on the 
flux given in the 2XMMi-DR3 catalogue. 
Cols.~[1] and [2] report the cluster identification number (detection Id, detid) 
and its name (IAUNAME), cols.~[3] and [4] provide the coordinates of X-ray 
emission in equinox J2000.0. The remaining columns are col.~[5] the XMM-Newton 
observation Id (obsid), col.~[6] the cluster spectroscopic redshift, col.~[7] 
the scale at the cluster redshift in kpc/$''$, col.~[8] the $R_{500}$ in kpc, 
cols.~[9] and [10] the 2XMMi-DR3 X-ray flux $F_{\rm cat}$ [0.5-2.0] keV and 
its error in units of $10^{-14}$\ erg\ cm$^{-2}$\ s$^{-1}$, cols.~[11] and 
[12] the estimated X-ray luminosity $L_{\rm cat}$ [0.5-2.0] keV and its error 
in units of $10^{42}$\ erg\ s$^{-1}$, cols.~[13] and [14] the cluster bolometric 
luminosity $L_{500}$ and its error in units of $10^{42}$\ erg\ s$^{-1}$, 
cols.~[15] and [16] the cluster mass $M_{500}$ and its error in units of 
$10^{13}$\ M$_\odot$, col.~[17] the {\tt objid} of the probable BCG in SDSS-DR10, 
cols.~[18] and [19] the BCG coordinates in equinox J2000.0, col.~[20] the 
apparent model magnitude m$_{\rm r}$ of the BCG, col.~[21] and [22] the weighted 
average spectroscopic redshift $\bar z_{\rm s}$ and the number of cluster 
members $N_{z_{s}}$ within 500 kpc with available spectroscopic redshifts 
that were used to compute the average redshift, col.~[23] and [24] the 
weighted average photometric redshift $\bar z_{\rm p}$ and the number of 
identified cluster member candidates $N_{z_{p}}$ within 500 kpc based on 
their photometric redshifts, col.~[25] the projected separation between the 
cluster X-ray position and the BCG position, cols.~[26] the NED name of 
previously known clusters in the literature, and col.~[27] a note indicating 
the object status, $``$Paper III$``$ refers to a new system confirmed using the 
current procedure, and $``$Paper II$``$ refers to a previously published 
system in Paper II with only photometric redshift and it is spectroscopically 
confirmed in the current work.


\subsection{Comparison of the derived X-ray parameters}

\begin{figure}
  \resizebox{\hsize}{!}{\includegraphics{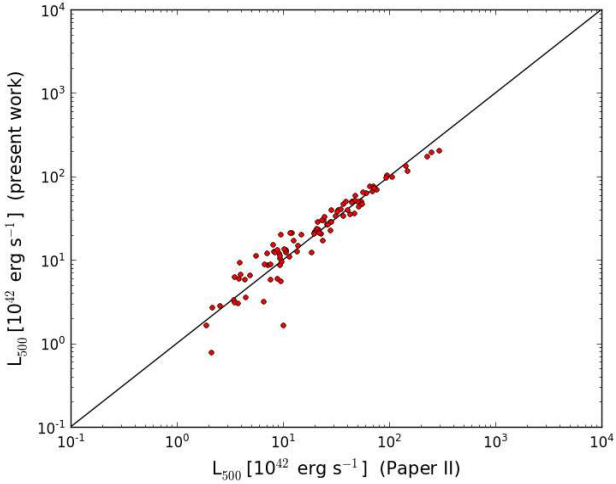}}
  \caption{Present measurements of $L_{500}$ plotted against the corresponding 
values from Paper II for 98 common clusters with fluxes obtained from the 
2XMMi-DR3 catalogue. The solid line represents the one-to-one relationship.} 
  \label{f:L500_L500PII}
\end{figure}

Among the clusters (316 systems) in common between the current cluster sample 
and that in Paper II, there were 98 clusters published in Paper II with 
their X-ray properties based on the flux given in the 2XMMi-DR3 catalogue.
Since we here used a different method to determine the X-ray bolometric $L_{500}$ 
from that used in Paper II, 
we present a comparison of the derived luminosities from the two 
methods. As shown in Fig.~\ref{f:L500_L500PII}, both methods give consistent 
values of $L_{500}$. The only two outliers are at the lowest redshifts 
(0.05 and 0.11) of the common sample. The median value of the ratios between 
the present $L_{500}$ estimates and the values from Paper II is 1.1. 
This offset of the current estimates of $L_{500}$ (10 percent) might 
arise from using two different procedures, the slight difference in the 
measured redshifts of the common systems, in addition to the two outliers 
mentioned above.

\begin{figure}
  \resizebox{\hsize}{!}{\includegraphics{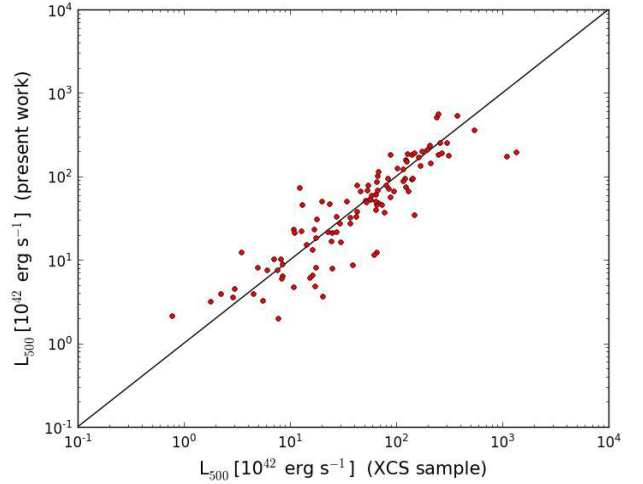}}
  \caption{Comparison of the present estimates of $L_{500}$ and the values  
from the XCS project for 107 clusters in common. The solid line shows the 
one-to-one relationship.} 
  \label{f:L500-Lxcs}
\end{figure}

Figure~\ref{f:L500-Lxcs} shows the good agreement between the current $L_{500}$ 
estimates and the corresponding values from the XCS project for 107 common 
objects. The ratio between the present $L_{500}$ to the corresponding values 
in the XCS project has a median of 0.95. 
Of the common sample, 33/107 have only photometric redshifts in the XCS 
sample, therefore our cluster sample provides spectroscopic confirmation for 
these systems. In general, there is consistency between the redshift 
measurements of the common sample apart from ten systems with 
$|z_{\rm present} - z_{\rm XCS}| > 0.05$, which have only photometric redshifts 
in the XCS sample. The mean and standard deviation of the redshift differences 
($z_{\rm present} - z_{\rm XCS}$) are 0.01 and 0.03, respectively.

\begin{figure}
  \resizebox{\hsize}{!}{\includegraphics{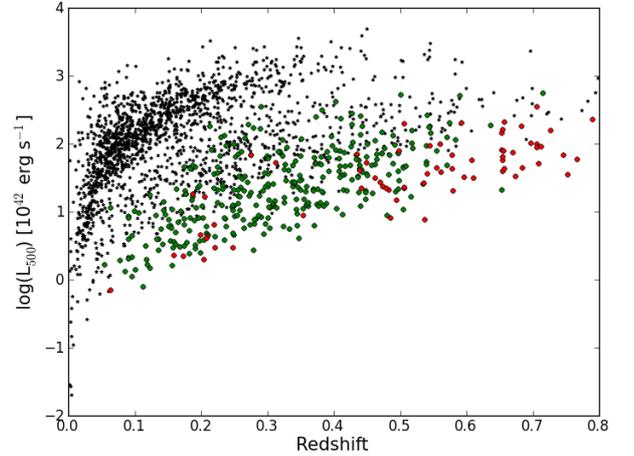}}
  \caption{Distribution of the X-ray bolometric luminosities, $L_{500}$, with 
redshift of the new cluster sample (red dots) from the current procedure,
the cluster sample in common with Paper II (green dots), and a sample of 1730 
clusters (black stars) below redshift 0.8 detected from ROSAT data 
\citep{Piffaretti11}.} 
  \label{f:L500-z-DR10}
\end{figure}

\begin{figure}
  \resizebox{\hsize}{!}{\includegraphics{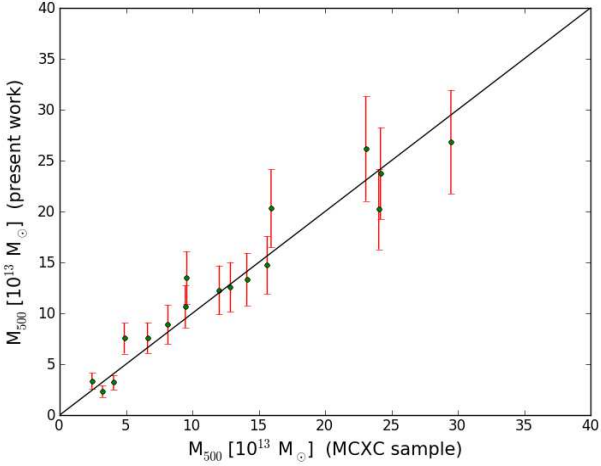}}
  \caption{Measurements of $M_{500}$ from our survey plotted against values 
from the MCXC catalogue for 17 clusters in common with ours. The solid line 
shows the one-to-one relationship.} 
  \label{f:M500-M500mcxc}
\end{figure}

The mass range of the cluster sample is $M_{500} \sim 2 - 28 \times 10^{13}$\ M$_\odot$ 
and the bolometric luminosity range is $L_{500} \sim 1 - 600 \times 10^{42}$\ erg\ s$^{-1}$.
Figure~\ref{f:L500-z-DR10} shows the distributions of $L_{500}$ as a function 
of redshift for the new cluster sample (67 systems) and the objects in common with
Paper II (316 objects) as well as for 1730 clusters ($z < 0.8$) from the MCXC 
catalogue, which was compiled from published X-ray selected cluster 
catalogues from ROSAT data \citep{Piffaretti11}.  Owing to the sensitivity 
of XMM-Newton and deeper exposures for some fields, the current cluster 
sample mostly includes low-luminosity groups and clusters at each redshift, as 
shown in Fig.~\ref{f:L500-z-DR10}.  
Cross-matching our cluster sample with the MCXC catalogue within a radius of 
30 arcsec yielded 17 clusters. The current estimates of $M_{500}$ agree 
well with the corresponding values in the MCXC catalogue of the common 
systems, as shown in Fig.~\ref{f:M500-M500mcxc}. The small overlap between 
the two samples is caused by the relatively small survey area in our project 
and by constraining our survey to include only systems identified as 
serendipitously detected sources and not targets in XMM-Newton observations.

In the new sample that comprises 67 X-ray selected clusters, only 12 
systems were previously known as X-ray detected clusters. They were  
published by \citet{Mehrtens12}, \citet{Clerc12}, \citet{Piffaretti11}, or 
\citet{Finoguenov07}. In total, about 25 percent of this new sample 
were previously known either in optical or X-ray band, the remainder 
of this sample are new systems. 

In Paper II, we presented the \ltr relation based on a cluster sample comprising 
345 systems in the redshift range $0.03 < z < 0.67$. In the future, we plan 
to extend the \ltr relation to include distant clusters up to slightly higher 
redshifts from the newly constructed cluster sample in this work. 
Since the current cluster 
sample provides spectroscopic confirmation for 78 clusters of the cluster 
sample published in paper II, we can re-measure their X-ray parameters with 
a better accuracy, which might help in reducing the scatter in the \ltr 
relation. We also plan to investigate the correlations between the X-ray 
and optical properties of the full cluster sample from our cluster survey 
that are optically confirmed so far.



\section{Summary}

We presented a sample of 383 X-ray selected galaxy groups and clusters 
associated with at least one LRG, which has a spectroscopic 
redshift in the SDSS-DR10. The redshifts of the associated LRGs were used to 
identify BCGs and other cluster galaxies with spectroscopic redshifts. 
The cluster spectroscopic redshift was computed as the weighted average of the 
available spectroscopic redshifts of the cluster galaxies within 500 kpc from 
the X-ray emission peak. The cluster sample spans a wide redshift range of 
$0.05 \le z \le 0.79$ with a median of $z = 0.34$.  Of the cluster 
sample, 264 are previously known as optically selected galaxy clusters. 
In addition to re-identifying and confirming the redshift estimates of 316 
clusters in common with the published cluster sample from our survey, we 
extended the optically confirmed cluster sample by 67 objects. Of this 
new sample that comprises 67 systems, about 75 percent are newly discovered 
groups and clusters, and about 80 percent are new X-ray detected clusters. 
Comparing the BCGs of the cluster sample with the colour and magnitude cuts 
of LRGs in the BOSS survey yielded that 92 percent of the BCGs are considered LRGs.
However, this percentage is dependent on the selection of the 
spectroscopically targeted LRGs in SDSS as well as on the current cluster 
identification procedure.    

The measured redshift and the X-ray flux in 0.5-2.0 keV given in the 2XMMi-DR3 
catalogue were used to determine the X-ray luminosity in 0.5-2.0 keV of the 
cluster sample. We converted the X-ray luminosity in 0.5-2.0 keV into bolometric
luminosity $L_{500}$ based on the available properties of the published cluster 
sample with X-ray spectroscopic parameters from our survey. 
This conversion yielded a scaling relation, which could be used to derive 
bolometric $L_{500}$ from the luminosity (0.5-2.0 keV) that is computed 
based on the $\beta$ model flux (0.5-2.0 keV) given in the 2XMMi-DR3 
catalogue.     

We also derived X-ray-luminosity-based masses of the cluster sample 
based on published scaling relation in the literature. 
Comparing the current estimates of the X-ray bolometric 
luminosity, $L_{500}$, with the available values from the XCS project, we 
found a good agreement between the two measurements. The distribution of 
X-ray luminosities of our cluster sample and ROSAT clusters with redshifts 
showed that we detected less-luminous groups and clusters at each redshift 
interval, and added a few tens of clusters at high redshifts.


 
\begin{acknowledgements}
This work is supported by the Egyptian Ministry of Higher Education and 
Scientific Research (MHESR) in cooperation with the Leibniz-Institut 
f{\"u}r Astrophysik Potsdam (AIP), Germany. We also acknowledge the financial 
support from the ARCHES project (7th Framework of the European Union, 
n$^{\circ}$ 313146) and the support by the Deutsches Zentrum f{\"u}r Luft- 
und Raumfahrt (DLR) under contract number  50 QR 0802. This work is based 
on observations obtained with XMM-Newton, an ESA science mission with 
instruments and contributions directly funded by ESA Member States and 
the USA (NASA). This research has made use of the NASA/IPAC Extragalactic 
Database (NED) which is operated by the Jet Propulsion Laboratory, California 
Institute of Technology, under contract with the National Aeronautics and Space 
Administration.
Funding for SDSS-III has been provided by the Alfred P. Sloan 
Foundation, the Participating Institutions, the National Science Foundation,
and the U.S. Department of Energy. The SDSS-III web site is 
http://www.sdss3.org/. 
SDSS-III is managed by the Astrophysical Research 
Consortium for the Participating Institutions of the SDSS-III Collaboration 
including the University of Arizona, the Brazilian Participation Group, 
Brookhaven National Laboratory, University of Cambridge, University of 
Florida, the French Participation Group, the German Participation Group, 
the Instituto de Astrofisica de Canarias, the Michigan State/Notre 
Dame/JINA Participation Group, Johns Hopkins University, Lawrence Berkeley 
National Laboratory, Max Planck Institute for Astrophysics, New Mexico State 
University, New York University, Ohio State University, Pennsylvania State 
University, University of Portsmouth, Princeton University, the Spanish 
Participation Group, University of Tokyo, University of Utah, Vanderbilt 
University, University of Virginia, University of Washington, and Yale 
University.
 \end{acknowledgements}



\begin{appendix}


\section{Selection criteria of LRGs}
We used the colour and magnitude cuts that were used to select spectroscopic  
targets to construct the BOSS galaxy sample in the the SDSS-III project. The  
selection criteria of galaxies targeted in BOSS were given in 
Padmanabhan et al. (2014) and were provided on the BOSS homepage
 \footnote{\url{http://www.sdss3.org/dr10/algorithms/boss_galaxy_ts.php}}.
The BOSS includes two samples of galaxies, one is the BOSS $``$LOWZ$``$ Galaxy Sample, $z \le 0.4$.
The selection cuts are as follows:

\begin{enumerate}

\item $|c_{\perp}| < 0.2$, to define the colour boundaries of the sample around 
a passive stellar population, where $c_{\perp} = (r - i) - (g - r)/4.0 - 0.18$. 

\item $r < 13.5 + c_{||}/0.3$, to select the brightest galaxies at each redshift, 
where $c_{||} = 0.7(g - r) + 1.2[(r - i) - 0.18)]$. 

\item $16 < r < 19.6$, to define the faint and bright limits.

\end{enumerate}

The other is the BOSS $``$CMASS$``$ Galaxy Sample, $0.4 < z < 0.8$.
The colour and magnitude cuts are as follows:

\begin{enumerate}

\item  $d_{\perp} > 0.55$, to isolate high-redshift objects, where 
       $d_{\perp} = (r-i) - (g-r)/8.0$.

\item  $i < 19.86+ 1.6(d_{\perp} - 0.8)$, to select the brightest or more 
       massive galaxies with redshift.
   
\item  $17.5 < i < 19.9 $, to define the faint and bright limits.

\item  $r - i < 2 $, to protect from some outliers.

\end{enumerate}

Note that we did not apply the criteria that were used to perform a star-galaxy
separation since we only considered objects that were classified as galaxies 
indicated by spectroscopic class parameters given in the {\tt SpecObj} table.
Note also that all colours were computed using model magnitudes while the magnitude 
cuts were applied on composite model (cmodel) magnitudes. All magnitudes were  
corrected for Galactic extinction following \cite{Schlegel98}.

\end{appendix}


 \bibliographystyle{aa}
 \bibliography{refbib_thesis}


\clearpage
\begin{landscape}
\begin{table}
\caption{\label{tbl:catalog_sample} The first ten entries of the new cluster 
sample (67 objects) from the current work in addition to a subsample 
(78 systems) that has only photometric redshifts in common with Paper II. 
The X-ray parameters are determined based on the flux given in the 
2XMMi-DR3 catalogue.}
{\footnotesize 
\begin{tabular}{c c c c c c c c c c c c c c c c }
 \hline
 \hline
  \multicolumn{1}{c}{detid\tablefootmark{a}} &
  \multicolumn{1}{c}{Name\tablefootmark{a}} &
  \multicolumn{1}{c}{ra\tablefootmark{a}} &
  \multicolumn{1}{c}{dec\tablefootmark{a}} &
  \multicolumn{1}{c}{obsid\tablefootmark{a}} &
  \multicolumn{1}{c}{z\tablefootmark{b}} &
  \multicolumn{1}{c}{scale} &
  \multicolumn{1}{c}{$R_{500}$} &
  \multicolumn{1}{c}{$F_{cat}$\tablefootmark{a,c}} &
  \multicolumn{1}{c}{$\pm eF_{cat}$} &
  \multicolumn{1}{c}{$L_{cat}$\tablefootmark{d}} &
  \multicolumn{1}{c}{$\pm eL_{cat}$} &
  \multicolumn{1}{c}{$L_{500}$\tablefootmark{e}} &
  \multicolumn{1}{c}{$\pm eL_{500}$} &
  \multicolumn{1}{c}{$M_{500}$\tablefootmark{f}} &
  \multicolumn{1}{c}{$\pm eM_{500}$} \\

  \multicolumn{1}{c}{} &
  \multicolumn{1}{c}{IAUNAME} &
  \multicolumn{1}{c}{(deg)} &
  \multicolumn{1}{c}{(deg)} &
  \multicolumn{1}{c}{} &
  \multicolumn{1}{c}{} &
  \multicolumn{1}{c}{kpc/$''$} &
  \multicolumn{1}{c}{(kpc)} &
  \multicolumn{1}{c}{} &
  \multicolumn{1}{c}{} &
  \multicolumn{1}{c}{} &
  \multicolumn{1}{c}{} &
  \multicolumn{1}{c}{} &
  \multicolumn{1}{c}{} &
  \multicolumn{1}{c}{} &
  \multicolumn{1}{c}{}  \\

(1) &  (2)  &  (3)  & (4)  &   (5)   &  (6)  &   (7)  &   (8)  &  (9)  &  (10)  &  (11) &  (12) & (13) & (14) & (15) &  (16)   \\
\hline 

005735 &     2XMM J003840.4+004746 &   9.66841 &   0.79636 & 0203690101 & 0.5553 & 6.44 &  542.28 &   1.44 & 0.18 &  17.86 &  2.20 &   44.88 &   5.02 &  8.31 &  1.69 \\
007554 &     2XMM J004304.2-092801 &  10.76751 &  -9.46695 & 0065140201 & 0.1866 & 3.12 &  585.47 &   6.84 & 1.06 &   6.74 &  1.05 &   18.50 &   2.61 &  6.87 &  1.45 \\
010986 &     2XMM J005556.9+003806 &  13.98720 &   0.63507 & 0303110401 & 0.2047 & 3.36 &  570.81 &   5.01 & 0.95 &   6.06 &  1.15 &   16.81 &   2.91 &  6.49 &  1.41 \\
016221 &     2XMM J012341.3+072323 &  20.92215 &   7.38985 & 0300000301 & 0.3418 & 4.86 &  640.75 &   5.90 & 0.85 &  23.02 &  3.33 &   56.53 &   7.43 & 10.68 &  2.17 \\
021043 &     2XMM J015558.5+053159 &  28.99394 &   5.53329 & 0153030701 & 0.4499 & 5.76 &  658.59 &   5.82 & 0.85 &  43.43 &  6.37 &  100.68 &  13.43 & 13.14 &  2.64 \\
021597 &     2XMM J020019.2+001931 &  30.08012 &   0.32553 & 0101640201 & 0.6825 & 7.07 &  629.03 &   4.17 & 0.52 &  85.10 & 10.59 &  185.57 &  20.98 & 15.09 &  2.98 \\
030746 &     2XMM J023346.9-085054 &  38.44543 &  -8.84844 & 0150470601 & 0.2653 & 4.08 &  615.99 &   5.95 & 1.09 &  12.93 &  2.36 &   33.47 &   5.55 &  8.71 &  1.84 \\
030889 &     2XMM J023458.7-085055 &  38.74463 &  -8.84868 & 0150470601 & 0.2590 & 4.01 &  581.98 &   4.15 & 0.53 &   8.54 &  1.09 &   22.96 &   2.67 &  7.29 &  1.51 \\
089821 &     2XMM J083114.4+523447 & 127.81014 &  52.57993 & 0092800201 & 0.6107 & 6.74 &  495.95 &   0.79 & 0.11 &  12.25 &  1.64 &   31.86 &   3.88 &  6.79 &  1.42 \\
089885 &     2XMM J083146.1+525056 & 127.94516 &  52.84719 & 0092800201 & 0.5190 & 6.23 &  616.14 &   3.50 & 0.21 &  36.78 &  2.24 &   86.56 &   4.80 & 11.67 &  2.26 \\

\hline
\end{tabular}
}
\end{table}
%

\addtocounter{table}{-1}
\begin{table}
\caption{\label{} continued.}
{\footnotesize  
\begin{tabular}{c c c c c c c c c c c c }
\hline
\hline
  \multicolumn{1}{c}{detid\tablefootmark{a}} &
  \multicolumn{1}{c}{objid\tablefootmark{g}} &
  \multicolumn{1}{c}{RA\tablefootmark{g}} &
  \multicolumn{1}{c}{DEC\tablefootmark{g}} &
  \multicolumn{1}{c}{$m_{r}$\tablefootmark{g}} &
  \multicolumn{1}{c}{$\bar z_{\rm s}$\tablefootmark{g}} &
  \multicolumn{1}{c}{$N_{z_{s}}$\tablefootmark{g}} &
  \multicolumn{1}{c}{$\bar z_{\rm p}$\tablefootmark{g}} &
  \multicolumn{1}{c}{$N_{z_{p}}$\tablefootmark{g}} &
  \multicolumn{1}{c}{offset\tablefootmark{g}} &
  \multicolumn{1}{c}{NED-Name} &
  \multicolumn{1}{c}{note\tablefootmark{h}}  \\
  
  \multicolumn{1}{c}{}&
  \multicolumn{1}{c}{(BCG)} &
  \multicolumn{1}{c}{(deg)} &
  \multicolumn{1}{c}{(deg)} &
  \multicolumn{1}{c}{(BCG)} &
  \multicolumn{1}{c}{} &
  \multicolumn{1}{c}{} &
  \multicolumn{1}{c}{} &
  \multicolumn{1}{c}{} &
  \multicolumn{1}{c}{(kpc)} &
  \multicolumn{1}{c}{} &
  \multicolumn{1}{c}{}  \\

  (1)  &  (17)  & (18)  &  (19)  &  (20)   &  (21) &  (22)  &   (23)   &  (24) & (25) & (26) & (27)   \\
\hline

005735 & 1237663204918428144 &   9.68054 &   0.78241 & 20.047 & 0.5553 &  3 & 0.5127 &  7 & 429.63 &                                   - & Paper-III \\
007554 & 1237652630713860232 &  10.80832 &  -9.47863 & 17.213 & 0.1866 &  2 & 0.1794 & 18 & 473.45 &                                   - & Paper-III \\
010986 & 1237663784740388918 &  14.02537 &   0.62659 & 17.537 & 0.2047 &  3 & 0.1951 & 20 & 473.61 &                                   - & Paper-III \\
016221 & 1237669767089357103 &  20.92160 &   7.39115 & 18.557 & 0.3418 &  1 & 0.3327 & 14 &  24.64 &                                   - & Paper-II \\
021043 & 1237678663047250389 &  28.98754 &   5.53072 & 19.553 & 0.4499 &  1 & 0.4258 & 12 & 142.32 &                                   - & Paper-II \\
021597 & 1237657071160263439 &  30.08100 &   0.32491 & 20.448 & 0.6825 &  1 & 0.6555 &  3 &  27.36 &                          SEXCLAS 03 & Paper-III \\
030746 & 1237653500970139807 &  38.44673 &  -8.84925 & 17.540 & 0.2653 &  1 & 0.2547 & 17 &  22.30 &                WHL J023347.2-085057 & Paper-II \\
030889 & 1237653500970270877 &  38.74547 &  -8.84926 & 17.762 & 0.2590 &  2 & 0.2528 & 17 &  14.70 &                                   - & Paper-II \\
089821 & 1237651701914141241 & 127.80965 &  52.57912 & 20.467 & 0.6107 &  1 & 0.6465 &  4 &  20.97 &                                   - & Paper-III \\
089885 & 1237651272960967114 & 127.94343 &  52.84937 & 19.251 & 0.5190 &  2 & 0.5165 & 13 &  54.28 &                WHL J083146.4+525057 & Paper-II \\

\hline
\end{tabular}
}

\tablefoot{The entire cluster catalogue is available online at CDS.
\tablefoottext{a}{Parameters extracted from the 2XMMi-DR3 catalogue.} 
\tablefoottext{b}{Spectroscopic redshift as given in col.~(21).}      
\tablefoottext{c}{2XMMi-DR3 flux, $F_{cat}$ [0.5-2.0] keV, and its errors in units of $10^{-14}$\ erg\ cm$^{-2}$\ s$^{-1}$.} 
\tablefoottext{d}{X-ray luminosity, $L_{cat}$ [0.5-2.0] keV, and its errors in units of $10^{42}$\ erg\ s$^{-1}$.}  
\tablefoottext{e}{X-ray bolometric luminosity at $R_{500}$, $L_{500}$ and its error in units of $10^{42}$\ erg\ s$^{-1}$.}  
\tablefoottext{f}{X-ray-luminosity-based mass at $R_{500}$, $M_{500}$ and its error in units of $10^{13}$\ M$_\odot$.} 
\tablefoottext{g}{Parameters obtained from the current detection algorithm in the optical band.} 
\tablefoottext{h}{A note about each system as $``$Paper III$``$: new cluster from the current work and $``$paper-II$``$: a cluster in Paper II and  confirmed spectroscopically with the present procedure.}
}

\end{table}
\end{landscape}



\end{document}